\documentclass{optica-article}

\journal{opticajournal} 

\articletype{Research Article}

\usepackage{lineno}
\usepackage{subcaption}
\usepackage{changes}
\usepackage{mdframed}
\definechangesauthor[name={del}, color=red]{del}
\definechangesauthor[name={add}, color=blue]{add}

\begin{document}
\begin{sloppypar}

\title{Research on fine co-focus adjustment method for segmented solar telescope}

\author{Kunyan Wang,\authormark{1,2} Yichun Dai,\authormark{1,3*} Bin Wang,\authormark{1,3} Xu Tan,\authormark{1,2} Dehua Yang,\authormark{4} and Zhenyu Jin\authormark{1,3}}

\address{\authormark{1}Yunnan Observatories, Chinese Academy of Sciences, Kunming 650216, China\\
\authormark{2}University of Chinese Academy of Sciences, Beijing 100049, China\\
\authormark{3}Yunnan Key Laboratory of Solar Physics and Space Science, 650216, China\\
\authormark{4}Nanjing Institute of Astronomical Optics and Technology, Chinese Academy of Sciences,  Nanjing 210042, China}

\email{\authormark{*}daiyichun@ynao.ac.cn} 


\begin{abstract*} 
For segmented telescopes, achieving fine co-focus adjustment is essential for realizing co-phase adjustment and maintenance, which involves adjusting the millimeter-scale piston between segments to fall within the capture range of the co-phase detection system. CGST proposes using a SHWFS for piston detection during the co-focus adjustment stage. However, the residual piston after adjustment exceeds the capture range of the broadband PSF phasing algorithm$(\pm 30 \mu m) $, and the multi-wavelength PSF algorithm requires even higher precision in co-focus adjustment. To improve the co-focus adjustment accuracy of CGST, a fine co-focus adjustment based on cross-calibration is proposed. This method utilizes a high-precision detector to calibrate and fit the measurements from the SHWFS, thereby reducing the impact of atmospheric turbulence and systematic errors on piston measurement accuracy during co-focus adjustment. Simulation results using CGST demonstrate that the proposed method significantly enhances adjustment accuracy compared to the SHWFS detection method. Additionally, the residual piston after fine co-focus adjustment using this method falls within the capture range of the multi-wavelength PSF algorithm. To verify the feasibility of this method, experiments were conducted on an 800mm ring segmented mirror system, successfully achieving fine co-focus adjustment where the remaining piston of all segments fell within $\pm 15 \mu m$.

\end{abstract*}

\section{Introduction}
The Chinese Giant Solar Telescope (CGST) is a next-generation giant solar telescope program jointly proposed by the Chinese solar physics community. A significant option for CGST's primary mirror involves employing ring-segmented mirrors. The scientific objective of CGST is to measure the delicate structures of magnetic and flow fields across various levels of the solar atmosphere, as well as their high spatial and temporal resolution evolutionary processes. For this purpose, the telescope is required to achieve co-phase within the 1 $\mu$ m wavelength range to realize high-resolution observation and research on the 20-km delicate structures of the solar surface\cite{1,2,3}. Due to the high precision requirement for co-phasing, which is in the order of nanometers, the capture range of the classical phasing algorithm is limited to ±30$\mu$m (Keck's broadband PSF phasing algorithm)\cite{4}. However, following the initial mechanical alignment, the piston between segments may reach the millimeter-scale. Therefore, it is necessary to adjust the large-scale piston before performing co-phasing, referred to as co-focus adjustment. In comparison with traditional co-focus adjustment, the paper focuses on fine co-focus adjustment, which not only entails adjusting the millimeter-scale piston to the capture range of the co-phase measurement system but also improving the accuracy of the co-focus adjustment, thereby simplifying the subsequent co-phase adjustment process.

Currently, the primary methods used for co-focus adjustment in segmented telescopes include spherometer measurement\cite{4}, Shack-Hartmann Wavefront Sensor(SHWFS) detection\cite{5,6}, and interferometer measurement\cite{7}. The 10m Keck telescope in the United States utilized a hand-held spherometer to adjust the large-scale piston within the capture range of the broadband PSF phasing algorithm $(\pm 30 \mu m) $\cite{4}. Similarly, telescopes like the 9.2m SALT in South Africa and the LAMOST in China employed SHWFS for co-focus adjustment, where defocus measurement reflects the segment's large-scale piston\cite{5,6,8,9}. However, the SHWFS of SALT had a practical detection accuracy of 60 $\mu$m(Root Mean Square, RMS) for large-scale piston, leading to the adoption of spherometer measurement, achieving co-focus adjustment with an accuracy of 15 $\mu$mRMS\cite{5}. Furthermore, interferometer measurement was proposed for co-focus adjustment in the 9.2m HET telescope in the United States, aiming for an accuracy of 25 $\mu$mRMS. However, due to insufficient robustness in the observing environment, SHWFS was ultimately employed for co-focus adjustment\cite{7}.

When the piston between segments reaches the millimeter-scale, it results in defocus. 
Therefore, the piston during the co-focus stage can be approximately obtained from the defocus measurement using a SHWFS. This method offers a more straightforward implementation than a spherometer or interferometer and can detect tip/tilt accurately. CGST plans to use SHWFS for piston measurement in the co-focus stage. However, due to atmospheric turbulence and systematic error\cite{10,11}, the current accuracy of SHWFS measurement for detecting large-scale piston is about 60 $\mu$mRMS (SALT), which falls short of the capture range of typical phasing algorithm ($\pm 30 \mu m$ for Keck's broadband PSF phasing algorithm). Additionally, the broadband PSF phasing algorithm is cumbersome because it requires scanning at a fixed step size, and the actuator displacements result in cumulative errors with too many scans. The co-focus measurement of TMT favors a multi-wavelength PSF detection method, which has a capture range of about $\pm 15 \mu m$, imposing higher demands on the accuracy of co-focus adjustment\cite{12,13}. This paper proposes a fine co-focus adjustment based on cross-calibration to improve the performance of the co-focus of CGST and facilitate the subsequent co-phase adjustment. The method employs a detector with higher detection accuracy to calibrate and fit the measurement results obtained from SHWFS. This approach aims to diminish the effects of atmospheric turbulence and systematic error on the measurement accuracy of the piston during the co-focus stage and improve the adjustment accuracy of the co-focus. To better verify the feasibility of this method, experiments were conducted on the 800mm ring segmented mirror system, and the adjustment results were analyzed.

Section 2 of this paper analyzes the fine co-focus adjustment method for CGST. It introduces the fine co-focus adjustment based on cross-calibration proposed in this paper and conducts a simulation analysis of the potential errors in the actual measurement. Section 3 presents the experimental results of fine co-focus adjustment using the above method on an 800mm ring segmented mirror system. The conclusion is to be presented in Section 4.

\section{Analysis of fine co-focus adjustment method for CGST}

An important alternative for the primary mirror of CGST is the utilization of ring-segmented mirrors. This configuration consists of 24 segments, each of which is an annular sector. The segments have a long base of 1040 mm, a short base of 779 mm, and a height of 1015 mm. The primary mirror is a parabolic reflector with a ring width of 1 m and a focal ratio of 1, as illustrated in Fig. 1\cite{1}. 

\begin{figure}[ht!]
    \centering
    \includegraphics[width=1\linewidth]{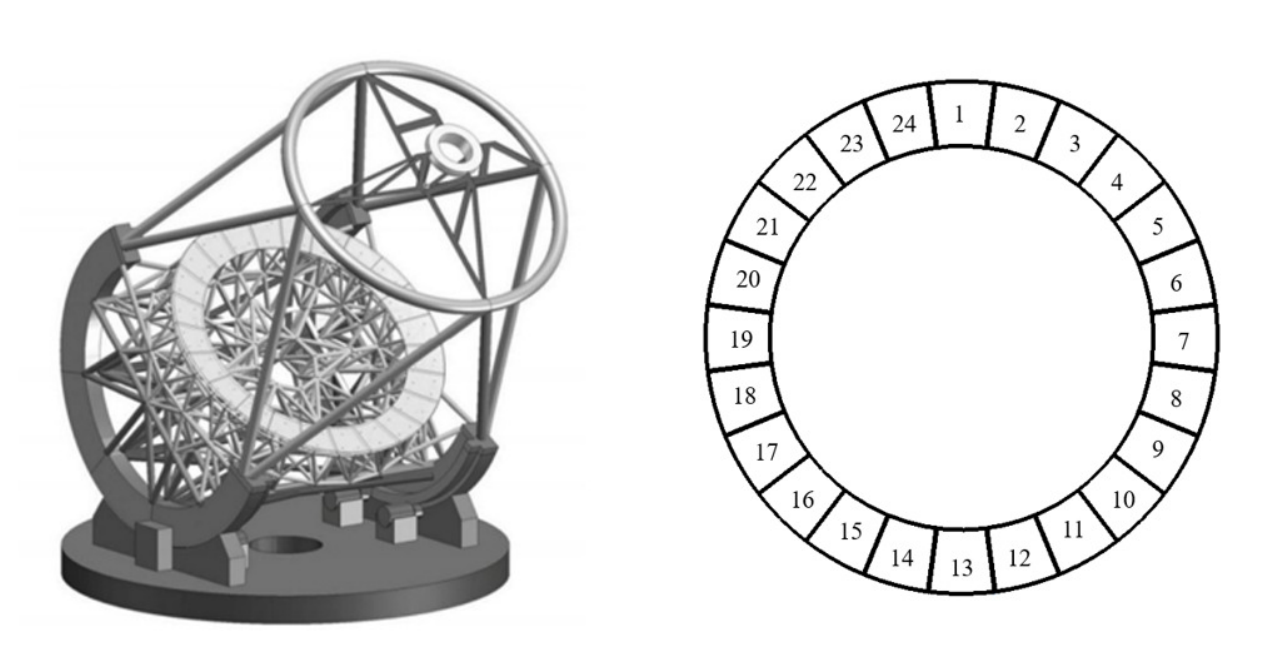}
    \caption{Schematic diagram of CGST and primary mirror}
    \label{fig1}
\end{figure}

\subsection{The principle and detection accuracy of SHWFS}
The SHWFS for co-focus adjustment is placed at the exit pupil of the optical system. Sixteen sub-apertures are planned to be allocated inside each segment to detect tip/tilt and piston during the co-focus stage. An additional two sub-apertures are placed at each edge for piston sensing during the co-phase stage, resulting in a total of 432 sub-apertures, as depicted in Fig. 2(a).

For SHWFS detection, $Z_2$, $Z_3$ and $Z_4$ denoting tip/tilt and defocus can typically be reconstructed using the modal method\cite{14,15}. During the co-focus stage, the piston between segments is large, which can be approximately obtained from defocus measurements. $Z_4$ denotes the piston during the co-focus stage in the paper. Due to the unique structure of the ring segmented mirrors, in this paper, the outer circle of the micro-lens array, corresponding to each segment, is selected as the unit-orthogonal circular domain for the Zernike polynomials, as illustrated in Fig.2(b). It has been verified that in the annular sector region, the 2nd Zernike polynomial is orthogonal to the 3rd Zernike polynomial, and the 3rd Zernike polynomial is orthogonal to the 4th Zernike polynomial, can be expressed as
\begin{equation}
\left\{\begin{array}{c}
\frac{1}{S} \iint_A(2 x)\cdot(2 y) d \sigma=0 \\
\frac{1}{S} \iint_A(2 x) \cdot \sqrt{3}\left(2\left(x^2+y^2\right)-1\right) d \sigma=0
\end{array}\right.
\label{eq:1}
\end{equation}
Where A denotes the annular sector region and S is the area of the annular sector. Therefore, using the modal wavefront reconstruction enables the decoupling of the segment's tip/tilt and piston.

\begin{figure}[ht!]
    \centering
    \begin{subfigure}[b]{0.4\textwidth}
      \includegraphics[width=\textwidth]{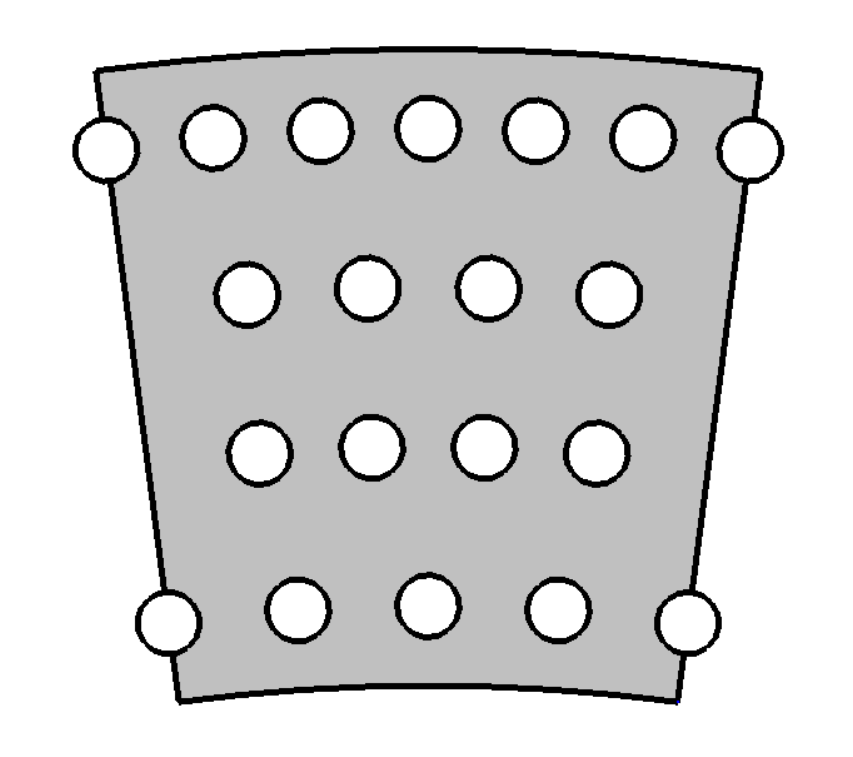}
      \caption{}
      \label{fig2a}
    \end{subfigure}%
    \begin{subfigure}[b]{0.45\textwidth}
      \includegraphics[width=\textwidth]{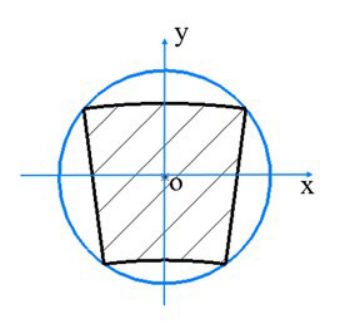}
      \caption{}
      \label{fig2b}
    \end{subfigure}
    \captionsetup{justification=centering}
    \caption{The SHWFS for CGST.(a)Micro-lens array corresponding to a segment. (b)The unit circle domain of the Zernike polynomials.}
\end{figure}

During actual detection, the accuracy of co-focus adjustment using SHWFS can be affected by environmental factors. In the presence of atmospheric turbulence, the mean-square value of the angle of arrival in the x or y direction on a circular aperture with a diameter D can be mathematically expressed as\cite{16} 
\begin{equation}
\left\langle\alpha_x^2\right\rangle=\left\langle\alpha_y^2\right\rangle=0.170\left(\frac{\lambda}{D}\right)^2\left(\frac{D}{r_0}\right)^{5 / 3}
\label{eq:2}
\end{equation}
where $\lambda$ is the wavelength and $r_0$ is the atmospheric coherence length. Suppose the sub-aperture of the micro-lens array corresponds to about 9.8 cm(the value of $r_0$ is approximately equivalent) on the primary mirror. In that case, the tilt error due to atmospheric turbulence is 0.55 arcsecond from Eq.(2) when $r_0$ is 10 cm and $\lambda$ is 650 nm. To mitigate the impact of atmospheric turbulence, integrating multiple frames of measurement data with successive short exposures over time can effectively suppress the effect\cite{17}. An atmospheric random phase screen based on the Kolmogorov model was generated to simulate atmospheric turbulence using the power spectral inversion method\cite{18}. The simulation results indicate that the accuracy of SHWFS for large-scale piston measurements is 195 $\mu$mRMS using a single frame and improves to 64 $\mu$mRMS when using 10 frames. 

Moreover, practical measurements with SHWFS may be affected by systematic error. For instance, the temperature gradient of the air in the measurement optical path and its slow change can introduce systematic measurement error. Compared to the random wavefront fluctuation caused by turbulence, the spatio-temporal frequency characteristic of the temperature gradient is lower, and the wavefront aberration is hardly to be eliminated by smoothing\cite{19}. This paper proposes a fine co-focus adjustment method based on cross-calibration to minimize the impact of various sources of uncertainty and improve the precision of co-focus adjustment.

\subsection{Fine co-focus adjustment based on cross-calibration}
\subsubsection {Basic Principles}
The principle of cross-calibration is to calibrate the measurement results of SHWFS using a more accurate measurement device. The cross-calibration is performed at multiple positions before and after the initial co-focus “0 point” obtained from SHWFS. The theoretical co-focus "0-point" position can be obtained more accurately by fitting the data. The mathematical expression of the cross-calibration method is shown in Eq. (3).
\begin{equation}
\hat{L}=f\left(Z_4\right)=\mathrm{k} * Z_4+b
\label{eq:3}
\end{equation}
In Eq. (3), L represents the detection value obtained from the calibration device with higher accuracy. $Z_4$ is the measured value of SHWFS, indicating piston during the co-focus stage. The intercept b in the equation represents the difference between the initial co-focus “0 point” obtained from SHWFS and the more accurate theoretical co-focus “0 point” obtained from the fine co-focus method. Therefore, after SHWFS detection and adjustment, the segment still needs to be moved by the amount of b.

The accuracy of the fine co-focus adjustment based on cross-calibration is determined by the solution precision of the intercept b, as indicated in Eq. (3). The standard deviation of b, denoted as  $s_b$, can be expressed as
\begin{equation}
s_b=\frac{s_L}{\sqrt{n}} \sqrt{1+\frac{{\overline{Z_4}}^2}{s_{Z_4}{ }^2}}
\label{eq:4}
\end{equation}
where n represents the number of measurement points; $s_L$ is the root mean square error (RMSE) of the fitting for L, given by 
\begin{equation}
s_L=\sqrt{\frac{1}{n-2} \sum_{i=1}^n\left(L_i-k * Z_{4,i}-b\right)^2}
\label{eq:5}
\end{equation}
$\overline{Z_{4}}$ represents the average of the measured values $Z_{4,i}$;$s_{Z_4}$
is the standard deviation of the measured value $Z_{4,i}$,given by 
\begin{equation}
s_{Z_4}=\sqrt{\frac{1}{n} \sum_{i=1}^n\left(Z_{4, \mathrm{i}}-\overline{Z_4}\right)^2}
\label{eq:6}
\end{equation}
Since the measurement points are situated before and after the initial co-focus “0 point” obtained from SHWFS,$\overline{Z_{4}}$ is small. Hence, Eq. (4) can be approximated as
\begin{equation}
s_b=\frac{s_L}{\sqrt{n}}
\label{eq:7}
\end{equation}
If the detection error of the calibration device is $\Delta_L$, the uncertainty of the intercept b can be expressed as shown in Eq. (8).

\begin{equation}
U_b=\sqrt{U_1{ }^2+{U_2}^2}=\sqrt{\left(t_{0.95}(n-2) * s_b\right)^2+\left(\Delta_L\right)^2}
\label{eq:8}
\end{equation}
In Eq. (8), $t_{0.95}(n-2)$ represents the t-distribution factor for a probability of 0.95 and n-2 degrees of freedom. This factor is approximately 1.8, and the exact value for different values of n can be obtained by referring to the appropriate table.

From Eq. (8), the accuracy of the fine co-focus adjustment based on cross-calibration is related to the number of measurement points n and the detection error of the calibration device $\Delta_L$. A higher co-focus adjustment accuracy can be achieved when n is larger and $\Delta_L$ is smaller. For instance, if the detection accuracy of SHWFS is 64 $\mu$mRMS, the value of $s_L$ is approximately 60 $\mu$m. When the number of measurement points n < 100, $U_1$ > 10 $\mu$m, $U_b$ is primarily determined by the magnitude of $U_1$, assuming the value of $\Delta_L$ is a few micrometers. However, when n surpasses 100, the influence of  $U_2$ in Eq. (8), which represents the detection accuracy of the calibration device, gradually becomes more significant as n increases. Theoretically, with a sufficiently large number of measurement points n, the highest detection accuracy of the fine co-focus adjustment based on cross-calibration is determined by the detection accuracy of the calibration device.

\subsubsection{Simulation analysis}
If we only consider the influence of atmospheric turbulence with an atmospheric coherence length $r_0 = 10 cm$, the simulation results of cross-calibration for measurement points of 10 and 50 are presented in Fig. 3. In the figure, the horizontal axis represents the $Z_4$ obtained from SHWFS by using 10 frames data for wavefront reconstruction. In contrast, the vertical axis indicates the detection value of the calibration device with a piston moving step of 20 $\mu$m. In the analysis of this paper, the uncertainty $U_b$ of the intercept b in Eq. (8) is employed to quantify the range of residual piston after fine co-focus adjustment. It can be inferred that the residual piston, after fine co-focus adjustment, can fall within the capture range of the multi-wavelength PSF algorithm when the number of measurement points n is approximately 50. Since the number of measurement points n utilized in the simulation analysis is less than 100, the detection error of the calibration device $\Delta_L$ can be neglected in the calculation. The range of residual pistons after adjustment using different measurement points are summarized in Table 1.The simulation results show that compared with SHWFS detection, the fine co-focus adjustment based on cross-calibration enhances the detection accuracy of large-scale piston. When using 10 frames data for wavefront reconstruction and employing 50 measurement points,  the range of residual piston after the fine co-focus adjustment is $\pm$14.2 $\mu$m.

\begin{figure}[ht!]
    \centering
    \includegraphics[width=1\linewidth]{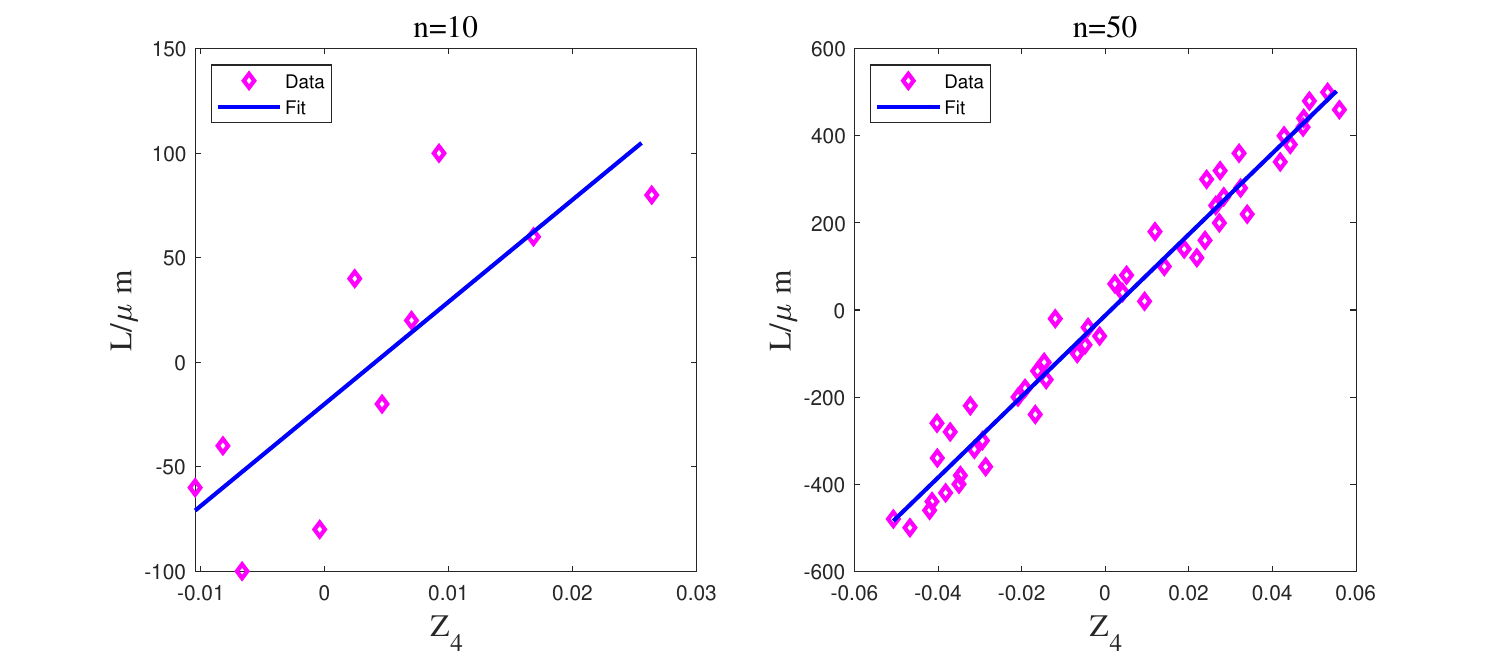}
    \caption{Simulation results of the fine co-focus adjustment based on cross-calibration (effect of atmospheric turbulence)}
    \label{fig3}
\end{figure}
  
\begin{table}[ht!]
\centering
\captionsetup{justification=centering}
\caption{The range of residual piston of the fine co-focus adjustment based on cross-calibration(effect of atmospheric turbulence)}
\label{tab:2}
\begin{tabular}{cccccc}
\hline  Number of measurement points& 10 & 20 & 30 &40  & 50\\
\hline The range of residual piston / $\mu$m &$\pm$29.7  &$\pm$22.1  & $\pm$17.1 &$\pm$15.5  & $\pm$14.2\\
 \hline 
\end{tabular}
\end{table}

Furthermore, in the presence of a systematic detection error $\sigma$, which may arise from factors such as temperature gradient or other sources, the result using cross-calibration is illustrated in Fig.4. The result indicates that the segment needs to be moved by -$\sigma$ after SHWFS detection and adjustment. Consequently, this adjustment method can also effectively mitigate the impact of systematic errors.

\begin{figure}[ht!]
    \centering
\includegraphics[width=0.6\linewidth]{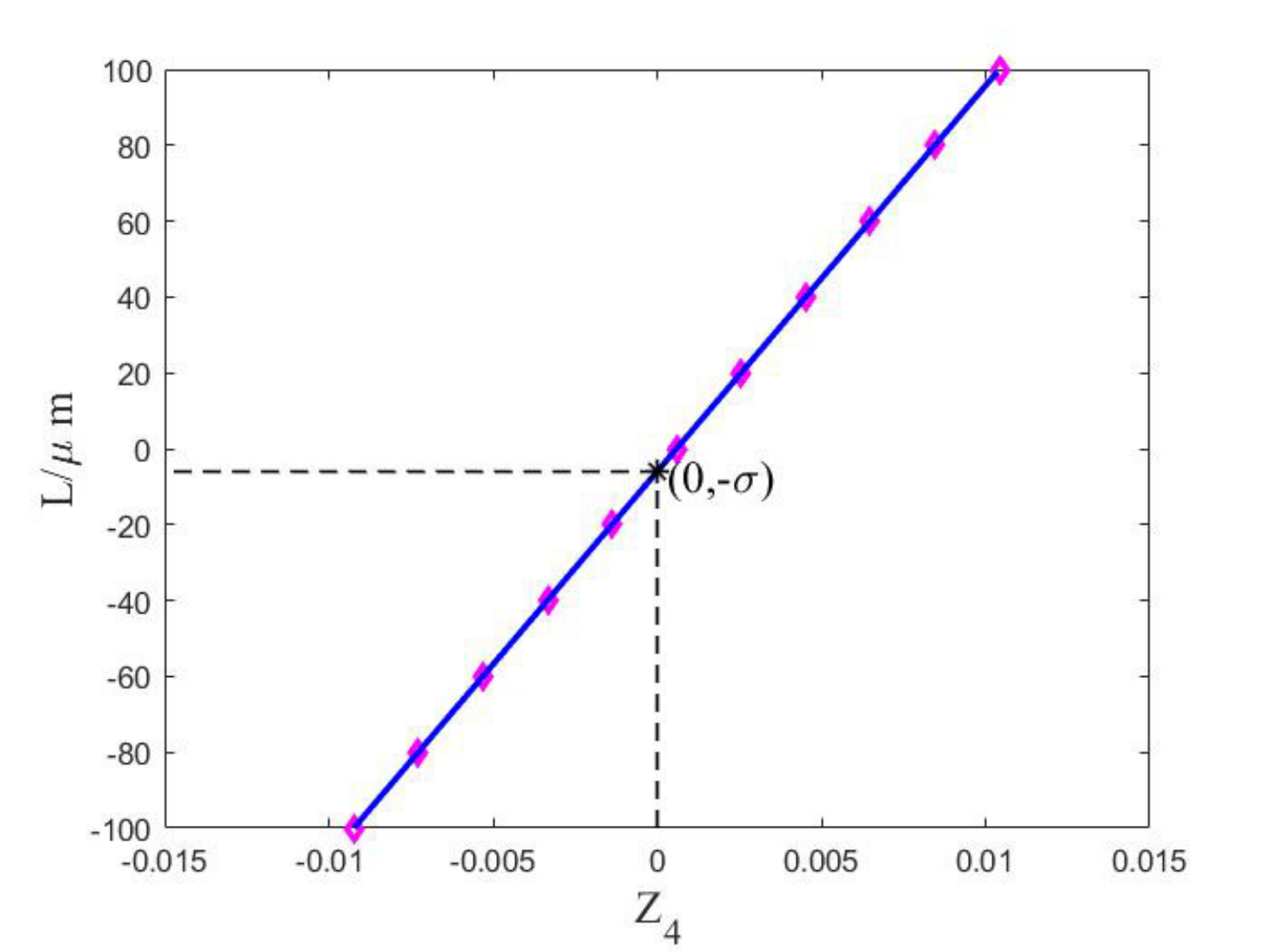}
    \caption{Simulation result of the fine co-focus adjustment based on cross-calibration(effect of systematic error)}
    \label{fig4}
\end{figure}

\section{Experimental results and analysis}
\subsection{Introduction to the experimental system}
The experimental system for fine co-focus calibration is a primary mirror composed of 8 annular sector spherical segmented mirrors, as illustrated in Fig. 5(a). The parameters of the segmented mirror are presented in Table 2. The co-focus detection optical path of the system is depicted in Fig. 6. In this configuration, a light source is positioned at the aplanatic points of the primary mirror, and SHWFS is employed to detect the tip/tilt and piston between the segments. A ring micro-lens array is situated at the primary mirror's exit pupil. Within the micro-lens array, each segment has seven internal sub-apertures and two sub-apertures at the edge, yielding 72 sub-apertures. During the co-focus adjustment stage, the seven internal sub-apertures are utilized to detect the segment's tip/tilt and piston, with the piston approximately obtained from defocus measurements. The two sub-apertures at the edge detect the piston during the co-phase adjustment stage. Once the co-focus adjustment is completed, a filter with a center wavelength of 636 nm and a bandwidth of 10 nm is employed for broadband scanning to adjust the piston, whose capture range is $\pm$15 $\mu$m. Therefore, the segment's fine co-focus adjustment accuracy must match the co-phase stage's capture range.

The calibration device is the LVDT (Linear Variable Differential Transformer), sold on the shelf with a detection accuracy of better than 3 $\mu$mRMS from the specification. It is mounted next to the actuator and is used to measure the actuator's linear displacement or displacement difference, as shown in Fig. 5(b).The actuator is screw-driven and capable of achieving high-precision micro-motions (during the co-phase stage). However, the actuator exhibits significant errors in the order of hundreds of micrometers for larger displacements, such as in co-focus adjustment. The LVDT measures the linear displacement of an object through the variation of the induced voltage in the secondary coil and behaves with high sensitivity and good linearity. In this paper's experiments, the displacement of the segment is determined by the readings of the LVDT. Additionally, the LVDT has a reference position known as the "0 point", which indicates that when the measured object is positioned at the center of the LVDT, the output voltage is zero. 

\begin{figure}[ht!]
    \centering
    \begin{subfigure}[b]{0.35\textwidth}
      \includegraphics[width=\textwidth]{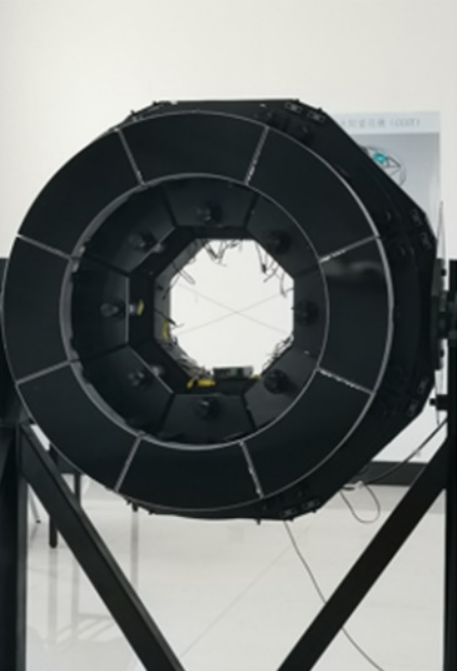}
      \caption{}
      \label{fig5a}
    \end{subfigure}%
    \begin{subfigure}[b]{0.58\textwidth}
      \includegraphics[width=\textwidth]{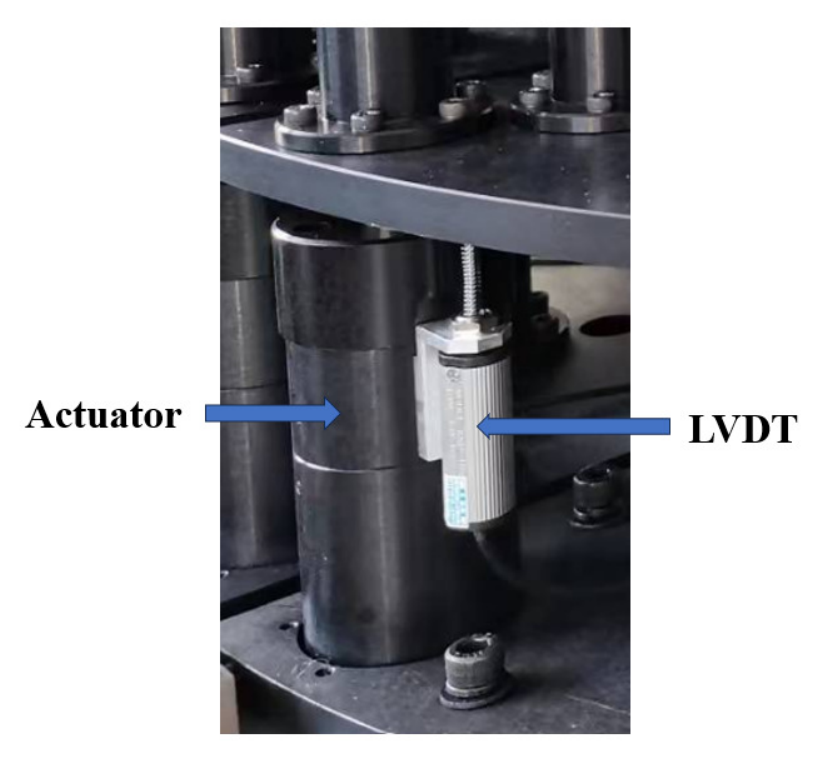}
      \caption{}
      \label{fig5b}
    \end{subfigure}
    \captionsetup{justification=centering}
    \caption{800mm ring segmented mirror system.(a)The primary mirror.(b)Actuator and LVDT}
\end{figure}

\begin{figure}[ht!]
    \centering
      \begin{minipage}{1\textwidth}
    \begin{subfigure}[b]{1\textwidth}
      \includegraphics[width=\textwidth]{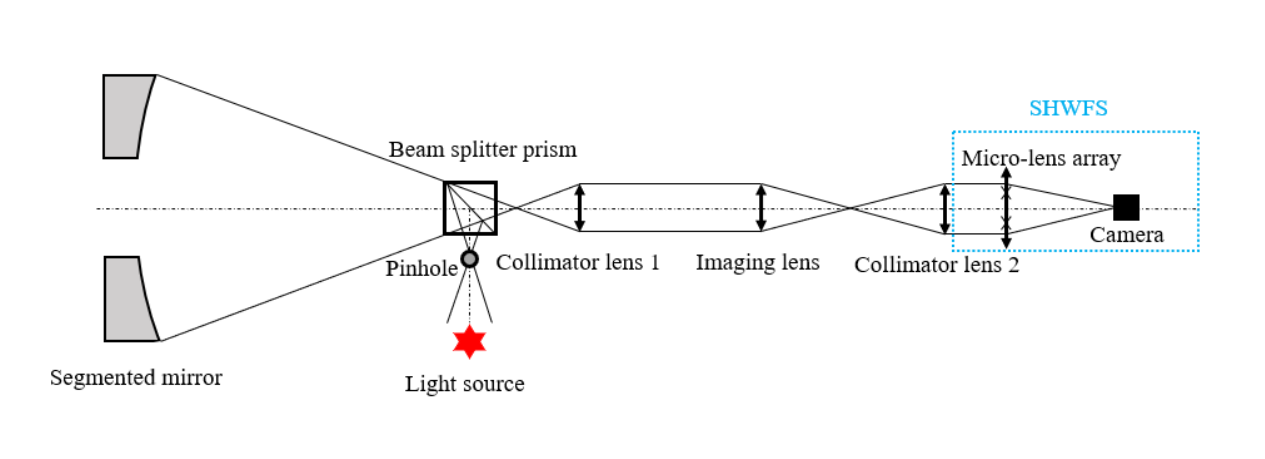}
      \caption{}
      \label{fig6a}
    \end{subfigure}%
    \end{minipage}\hfill
      \centering
       \begin{minipage}{1\textwidth}
         \centering
    \begin{subfigure}[b]{0.7\textwidth}
      \centering
      \includegraphics[width=\textwidth]{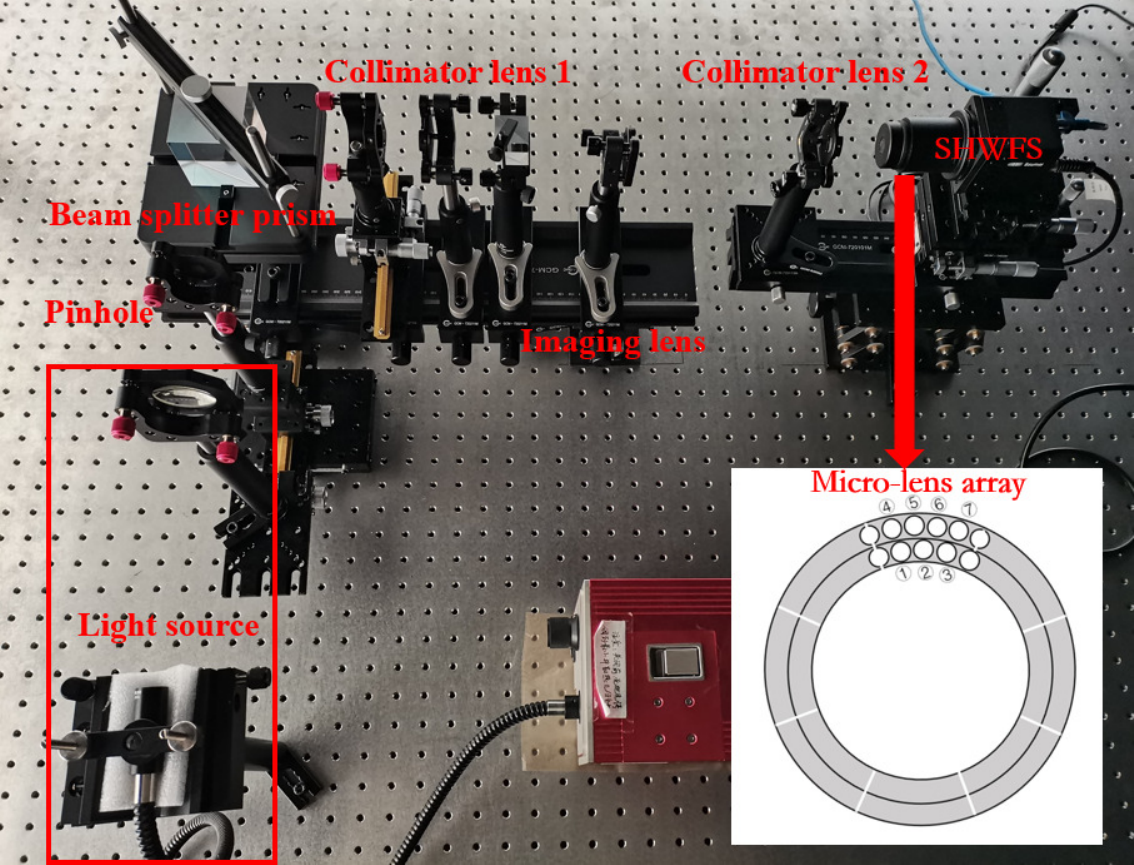}
      \caption{}
      \label{fig6b}
    \end{subfigure}
    \end{minipage}\hfill
    \captionsetup{justification=centering}
    \caption{Co-focus detection optical path.(a)Schematic diagram.(b)Physical image}
\end{figure}

\begin{table}[ht!]
\centering
\captionsetup{justification=centering}
\caption{Optical parameters of the 800mm ring segmented mirror system}
\label{tab:3}
\begin{tabular}{ccc}
\hline  Parameter& Symbol & Value\\
\hline Diameter of primary mirror&D/mm&800\\
Radius of curvature&R/mm&3200\\
Ring width&$\Delta$D/mm&120\\
Focal ratio&f/$\#$&2\\
Focal length of collimator lens 1&f1/mm &37.5\\
Focal length of collimator lens 2&f2/mm &72.38\\
Focal length of imaging lens&f3/mm &78.39\\
Diameter of pinhole& d1/ $\mu$m&20\\
Diameter of sub-aperture& d2/ $\mu$m&583\\
Focal length of micro-lens array& f4/mm&77\\
Detector resolution& p*p/(pixel*pixel)&2048*2048\\
Pixel size&d3*d3/( $\mu$m * $\mu$m) &5.5*5.5\\
 \hline 
\end{tabular}
\end{table}

\subsection{The detection accuracy of LVDT and SHWFS}
For our SHWFS, the detection accuracy of tip/tilt is higher than 0.085 arcsecond RMS when the centroid sensing obtains sub-pixel resolution. This level of accuracy corresponds to an actuator length of 0.08 $\mu$mRMS. The LVDT has a detection accuracy of better than 3 $\mu$mRMS from the specification. The actual detection accuracy of the LVDT should be determined before it is used as the high detection means for cross-calibration of SHWFS. Due to the high accuracy of the SHWFS in detecting tip/tilt, we utilized tip measurements to calibrate the measurement accuracy of the LVDT. First, a movement of 4 $\mu$m was applied to the controller of the actuator M1 to produce a tip, while the readings of the LVDT and the detected value $Z_3$ of the SHWFS were recorded. Subsequently, using the relationship between $Z_3$ and the displacement of actuator M1, $Z_3$ was converted into the corresponding actuator displacement, considering it as a reference value. The difference between the measured value of the LVDT and the reference value represented the detection error of the LVDT. Fig.7 illustrates the detection errors of the LVDT from 40 measurements, with an RMS value of 1.93 $\mu$m. Therefore, the LVDT can be utilized to cross-calibrate the SHWFS.

\begin{figure}[ht!]
    \centering
    \includegraphics[width=0.75\linewidth]{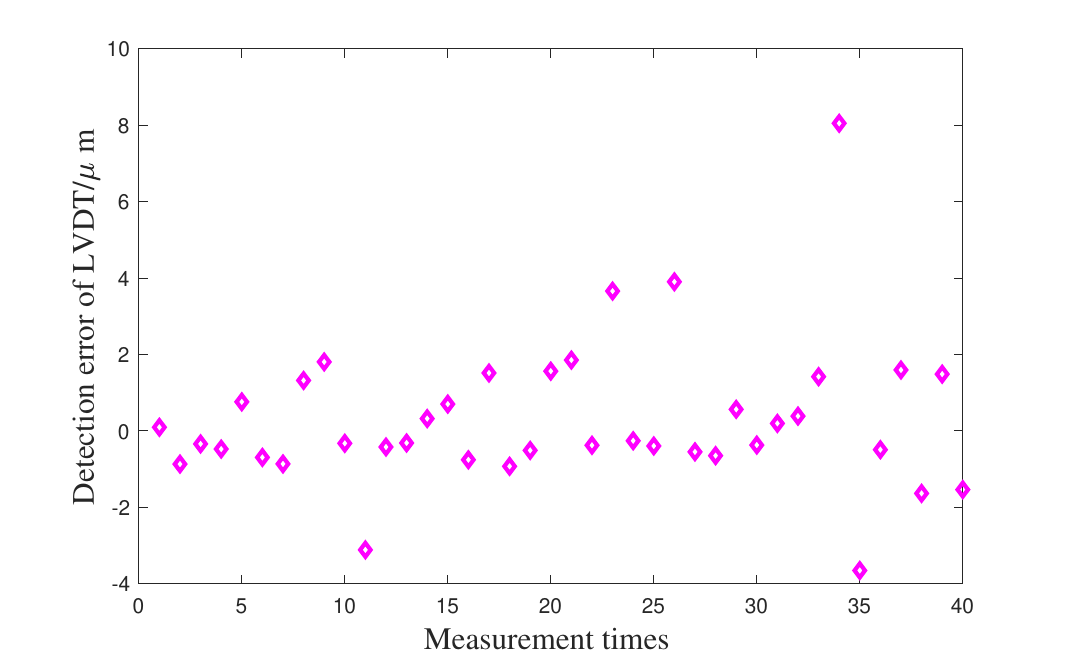}
    \caption{The detection error of the LVDT}
    \label{fig7}
\end{figure}

Similarly, the accuracy of LVDT is higher than SHWFS in detecting large-scale piston, so the measurement values obtained from the LVDT can be used to evaluate the accuracy of SHWFS in detecting large-scale piston. First, a displacement of 50 $\mu$m was applied as the input to the controllers of three actuators corresponding to the segment, generating a piston. Simultaneously, the readings of the LVDT and $Z_4$ from the SHWFS were recorded, with the LVDT readings as the reference value. Subsequently, using the relationship between $Z_4$ and the displacement of the actuators, $Z_4$ was converted into corresponding actuator displacement. The difference between the converted value and the reference value represented the detection error of the piston. Fig.8 illustrates the detection errors of 40 measurements of piston obtained from the SHWFS. The RMS value of the detection errors is 20.41 $\mu$m, while the peak-to-valley (PTV) value is 90.99 $\mu$m.

\begin{figure}[ht!]
    \centering
    \includegraphics[width=0.75\linewidth]{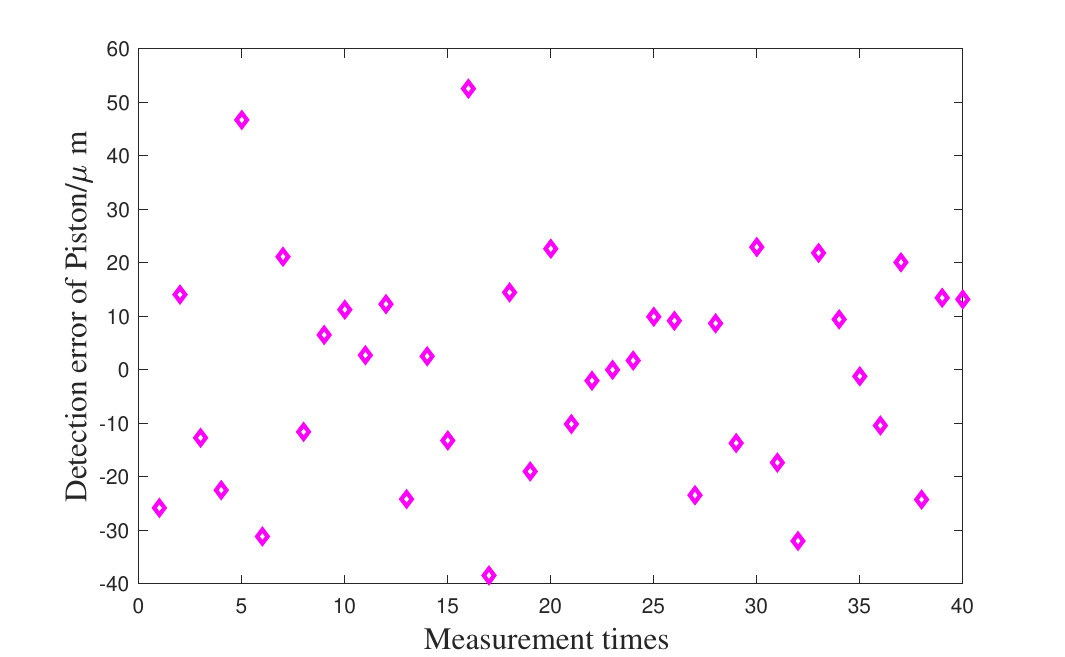}
    \caption{The detection error of the piston(SHWFS)}
    \label{fig8}
\end{figure}

\subsection{Adjustment results and analysis of fine co-focus adjustment based on cross-calibration}
Fine co-focus adjustment using LVDT and SFWFS was as follows: Firstly, the initial co-focus "0 point" was obtained after SHWFS detection and adjustment.To ensure that the residual piston after fine co-focus adjustment could fall into the capture range of the broadband PSF phasing of $\pm$ 15 $\mu$m, the number of measurement points, denoted as n, could be estimated using Eq. (8) to be 10, where the RMSE of the LVDT fitting $s_L$ was approximately 15 $\mu$m. Consequently, five points were measured in 50 $\mu$m(controller input value for actuator) increments both before and after the initial co-focus "0 point", and the corresponding LVDT readings and $Z_4$ measured by SHWFS were simultaneously recorded. A least-squares fitting was then performed on the variations in LVDT readings and the corresponding $Z_4$ .Finally, the value of LVDT when $Z_4$=0 was calculated, representing the additional displacement required for the segment to obtain a more precise theoretical co-focus “0-point”. Fig. 9 illustrates the fine co-focus adjustment results of segment 1 repeated four times. The horizontal axis represents the detection value $Z_4$ of the SHWFS, while the vertical axis represents the variation in LVDT readings. "-22$\mu$m", "-26$\mu$m", "-15$\mu$m", and "-30$\mu$m" represent the adjustments required for segment 1  after SHWFS detection and adjustment (results of 4 experiments).

\begin{figure}[ht!]
    \centering
    \includegraphics[width=1\linewidth]{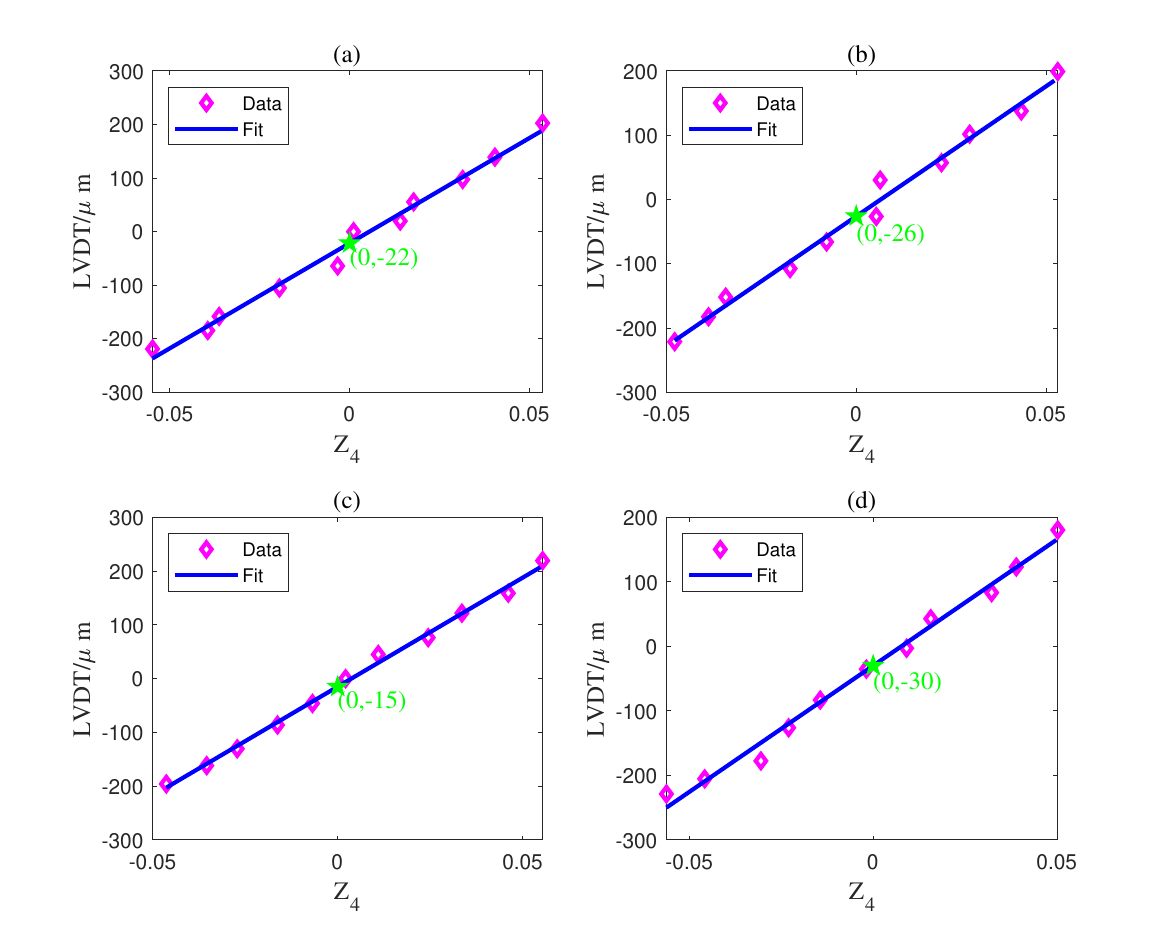}
    \caption{Results of multiple co-focus adjustments for segment 1.}
    \label{fig9}
\end{figure}

In the fine co-focus adjustment, the range of residual piston is estimated from the uncertainty $U_b$ of the intercept b in Eq. (8), where the value of $t_{0.95}(n-2)$ is determined to be 1.86 for n is 10. The standard deviation $s_b$ of the intercept b, obtained by fitting the measured data of the eight segments, is presented in Table 3. The detection error $\Delta_L$ of the LVDT is 3 $\mu$mRMS, which can be disregarded in the calculation. The range of residual pistons after the adjustment of the eight segments using the fine co-focus adjustment based on cross-calibration are displayed in Table 4.

\begin{table}[ht!]
\centering
\captionsetup{justification=centering}
\caption{Standard deviation of the intercept b obtained from real measurements of 800mm ring segmented mirror system}
\label{tab:4}
\begin{tabular}{ccccccccc}
\hline  segment&1 & 2&3 &4 &5 &6 &7&8\\
\hline $s_b$/ $\mu$m&6.1 &7.0&4.9 &4.6 & 7.0&2.7 &3.0&4.7\\
\hline
\end{tabular}
\end{table}

\begin{table}[ht!]
\centering
\captionsetup{justification=centering}
\caption{The range of residual piston after fine co-focus adjustment for 800mm ring segmented mirror system}
\label{tab:5}
\begin{tabular}{ccccccccc}
\hline  segment&1 & 2&3 &4 &5 &6 &7&8\\
\hline The range of residual piston / $\mu$m&$\pm$11.3 &$\pm$13.0&$\pm$9.1 &$\pm$8.6 & $\pm$13.0&$\pm$5.0 &$\pm$5.6&$\pm$8.7\\
\hline
\end{tabular}
\end{table}

Based on the experimental results presented in Table 4, it can be observed that when the number of measurement points is 10, the detection accuracy of fine co-focus adjustment based on cross-calibration can reach 26$\mu$m(PTV), and the residual pistons of all segments after fine co-focus adjustment fall into the capture range of the broadband PSF phasing($\pm$15 $\mu$m). Furthermore, these residual pistons also remain within the capture range of the multi-wavelength PSF algorithm. As a result, the subsequent co-phase adjustment can be carried out. 

\section{Conclusion}
This paper presents the fine co-focus adjustment based on cross-calibration with high precision for the CGST. The accuracy of this method relies on two key factors: the number of measurement points used for cross-calibration and the detection accuracy of the calibration device. This method effectively mitigates the detection error caused by atmospheric turbulence and reduces the impact of systematic error. Compared to the SHWFS detection system,  the method significantly improves the accuracy of co-focus adjustment. Moreover, when using 50 measurement points, the residual piston is within $\pm$14.2 $\mu$m, ensuring it falls within the capture range of the multi-wavelength PSF phasing algorithm. As a result, this method simplifies the following co-phase adjustment process.

Co-focus adjustment experiments were conducted on the 800mm ring segmented mirror system using the proposed method, which involved the cross-calibration of SHWFS with LVDT. The LVDT demonstrates a measurement accuracy better than 3 $\mu$mRMS. Compared to the SHWFS detection system, the proposed fine co-focus adjustment method(when using 10 measurement points) improves the co-focus adjustment accuracy of the experimental system from 91$\mu$m(PTV) to 26$\mu$m(PTV). The residual pistons for all segments fall within the capture range of the broadband PSF phasing algorithm($\pm$15 $\mu$m). Moreover, the LVDT has a "0 point" that can be used to record the segment's state during the co-focusing, which is beneficial for the segment to return to the co-focus state quickly. This fine co-focus adjustment method, employing LVDT and SHWFS cross-calibration, is also a suitable reference scheme for CGST's fine co-focus adjustment. Furthermore, the "0 point" of the LVDT may experience drift due to environmental conditions. Therefore, further experimental research is necessary to address this potential issue.

\begin{backmatter}
\bmsection{Funding}
 National Natural Science Foundation of China (12273109);Yunnan Province Young and Middle-aged Academic and Technical Leaders Reserve Talents Project (202405AC350004);Yunnan Key Laboratory of Solar Physics and Space Science(202205AG070009);Yunnan Revitalization Talent Support Program(202305AS350029, 202305AT350005);Yunnan Provincial Science and Technology Department(202103AD150013).

\bmsection{Acknowledgments}
We would like to express our heartfelt gratitude to the Laboratory of Astronomical Technologies of Yunnan Observatories, Chinese Academy of Sciences, for its strong support and assistance throughout this research. The laboratory's facilities and resources have been instrumental in ensuring the smooth progress of our study.

\bmsection{Disclosures}
The authors declare that there are no conflicts of interest related to this paper.

\bmsection{Data availability}  Data underlying the results presented in this paper may be obtained from the authors upon reasonable request.

\end{backmatter}







\end{sloppypar}
\end{document}